\documentclass[conference]{IEEEtran}

\usepackage[a-1b]{pdfx}

\usepackage{algorithm2e}
\usepackage{lipsum}
\usepackage{subcaption}
\usepackage[frozencache]{minted}
\usepackage{url}
\usepackage{xspace}
\usepackage{graphicx}
\usepackage{booktabs}
\usepackage{textcomp}
\usepackage{tikz}
\usepackage[para,online,flushleft]{threeparttable}
\usepackage{tabulary}
\usepackage{microtype}
\usepackage{tikz}
\usepackage[utf8]{inputenc}

\makeatletter
\renewcommand{\@endalgocfline}{\relax}
\makeatother
\SetKwProg{Procedure}{Procedure}{}{}

\usetikzlibrary{shapes, arrows, automata, positioning, fit, shadows, decorations.pathmorphing}
\usetikzlibrary{calc, backgrounds}

\def\hardgists{HG2.9k}

\newcommand\insight[1]{
	\noindent 
	\fcolorbox{gray!20}{gray!20}{
		\parbox{0.92\columnwidth}
		{#1}
		\hspace*{0.5ex}
	}
}

\begin{document}

    \title{DockerizeMe: Automatic Inference of Environment Dependencies for Python Code Snippets}

    \author{
        \IEEEauthorblockN{Eric Horton, Chris Parnin}
        \IEEEauthorblockA{
            NC State University\\
            Raleigh, NC, USA\\
            Email: \{ewhorton, cjparnin\}@ncsu.edu
        }
    }

    \maketitle

    \begin{abstract}

    Platforms like Stack Overflow and GitHub's gist system promote the sharing of ideas and programming techniques via the distribution of code snippets designed to illustrate particular tasks. Python, a popular and fast-growing programming language, sees heavy use on both sites, with nearly one million questions asked on Stack Overflow and 400 thousand public gists on GitHub. Unfortunately, around 75\% of the Python example code shared through these sites cannot be directly executed. When run in a clean environment, over 50\% of public Python gists fail due to an import error for a missing library.
    
    We present DockerizeMe, a technique for inferring the dependencies needed to execute a Python code snippet without import error. DockerizeMe starts with offline knowledge acquisition of the resources and dependencies for popular Python packages from the Python Package Index (PyPI). It then builds Docker specifications using a graph-based inference procedure. Our inference procedure resolves import errors in 892 out of nearly 3,000 gists from the Gistable dataset for which Gistable's baseline approach could not find and install all dependencies.
    
\end{abstract}


    \section{Introduction}{
    
    Sharing code snippets to illustrate a specific task is frequently used practice within the software engineering industry~\cite{Sillito:2012}. Due to the importance of sharing examples, platforms like Stack Overflow and GitHub's gist system have been created to facilitate social learning~\cite{Parnin:2012:crowd, Yang:2016, Ford:2016} through community driven interaction. GitHub gists are short but complete programs, often only a single file.

    Unfortunately, many code snippets do not contain information regarding the system configuration needed for proper execution, as system configuration is not an inherent property of code. Consider Sentry, an error reporting system. The official client for Python, Raven, supports the Flask framework~\cite{Sentry:Flask}. Examples\footnote{\url{https://gist.github.com/1cdd9646046ae10c2932}} demonstrating how to use Sentry with Flask often have no indication that it needs to be installed with Flask extras, causing developers to encounter runtime errors\footnote{\url{https://github.com/getsentry/raven-python/issues/1075}}.
    
    This is a wide-spread problem. Research by Yang et al. found that only 25\% of code snippets from Stack Overflow can be run without error~\cite{Yang:2016}. Horton and Parnin later found 24.4\% of Python gists from GitHub to run without error~\cite{Horton:Gistable:2018}. The main cause of failure, experienced by 52.4\% of the gists evaluated, was a dependency error. Seo et al. also found that approximately 50\% of build errors are caused by dependencies~\cite{Seo:2014:BuildErrors}. Further, the effort involved in manually constructing an environment specification is non-trivial --- developers can spend between 20 minutes and 2 hours creating a Dockerfile for a single code snippet, and often fail to construct a valid specification~\cite{Horton:Gistable:2018}. Common challenges include mapping a code resource to its originating package and determining the correct order of installation for transitive dependencies.
    
    This work focuses on automating the dependency resolution process of system configuration management for both language-level and system-level dependencies. Before performing dependency resolution, we build an offline knowledge base from two sources. First, we process existing packages on the Python Package Index (PyPI) by extracting declared resources and using dynamic analysis to determine possible dependencies. Second, we inspect project configuration files from GitHub and generate association rules for pairs of packages which are frequently seen together. Environment inference for a code snippet is performed by querying the knowledge base to map code resources to installable packages. A search algorithm then resolves all transitive dependencies in a consistent order.
    
    We implement DockerizeMe, a tool for applying dependency resolution to a code snippet and generating a Dockerfile for the corresponding environment. Unlike other approaches to automated software configuration, which focus on repairing configuration errors (\cite{Weiss:Tortoise:2017, Hassan:2018, Ren:2018}), DockerizeMe focuses on inferring a complete configuration without external inputs. Being able to automatically infer code dependencies has the potential to save developers time, reduce costs of learning and development, and enable repair and verification of code snippets in online platforms. It is also a first step towards fully automated software configuration management.

    To evaluate DockerizeMe, we performed environment configuration for 2,891 gists from the Gistable dataset which still failed due to Python's \texttt{ImportError} after applying Gistable's Environment Inference Algorithm (Section III-C from \cite{Horton:Gistable:2018}). DockerizeMe successfully removed import errors for 892 gists.
    
     \begin{figure*}[ht]
        
        \centering
        \begin{tabular}{c|c}
            \begin{subfigure}[b]{0.55\textwidth}
                \centering
                \begin{minted}[fontsize=\footnotesize,linenos,autogobble,breaklines]{python}
                    import pcapy
                    from impacket.ImpactDecoder import *
                    
                    # list all the network devices
                    pcapy.findalldevs()
                    
                    max_bytes = 1024
                    promiscuous = False
                    read_timeout = 100 # in milliseconds
                    pc = pcapy.open_live("eth0", max_bytes, promiscuous, read_timeout)
                    
                    pc.setfilter('tcp')
                    
                    # callback for received packets
                    def recv_pkts(hdr, data):
                        packet = EthDecoder().decode(data)
                        print packet
                    
                    packet_limit = -1 # infinite
                    pc.loop(packet_limit, recv_pkts) # capture packets
                \end{minted}
                \caption{snippet.py: \url{https://stackoverflow.com/a/4948251/8588856}}
                \label{fig:motivation:code}
            \end{subfigure}
            &
            \begin{subfigure}[b]{0.35\textwidth}
                \centering
                \begin{minipage}{.8\textwidth}
                    \begin{minted}[fontsize=\footnotesize,linenos,autogobble,breaklines]{Dockerfile}
                        FROM python:2.7.14

                        # Update APT package list and install required system dependency
                        RUN apt-get update
                        RUN apt-get install -y libpcap-dev
                        
                        # Install Python dependencies not included with snippet.py
                        RUN pip install pcapy
                        RUN pip install impacket
                        
                        COPY snippet.py /scripts/snippet.py
                        CMD python /scripts/snippet.py
                    \end{minted}
                \end{minipage}
                \caption{Dockerfile}
                \label{fig:motivation:Dockerfile}
            \end{subfigure}
        \end{tabular}
        \caption{
            (a) Code snippet for capturing packets on a network interface. (b) Dockerfile for running the code snippet.
        }
        \label{fig:motivation}
    \end{figure*} 

    In summary, this work contains the following contributions
    \begin{itemize}
        \item A technique for computing package dependencies using static analysis, dynamic analysis, and developer generated knowledge sources.
        \item An inference algorithm for direct and transitive dependencies that respects installation order.
        \item DockerizeMe: a tool for building inferred environments. (\url{https://github.com/dockerizeme/dockerizeme})
        \item An empirical evaluation of DockerizeMe's effectiveness and a categorization of additional challenges in environment configuration.
    \end{itemize}

}

    \section{Motivating Example}{
   
    Many code snippets shared as examples are not directly executable, often because they depend on external libraries that are not present by default on a developer's system~\cite{Yang:2016, Horton:Gistable:2018}. Discovering all required dependencies is a time consuming process that even experienced developers have trouble with~\cite{Horton:Gistable:2018}. Consider the following scenario:
    
    A developer is working on a networking component and needs to parse packets from a network interface. They would like to use Python, and search for Python libraries that can be used to perform this task. Fortunately, other developers have previously asked for recommendations on libraries to solve the same problem. They quickly come across a post on Stack Overflow\footnote{https://stackoverflow.com/questions/4948043/} with a couple different recommendations. 
    
    The accepted answer recommends Scapy, but it is licenced under GPLv2, a copyleft license which the developer cannot use for their project. However, another answer recommends Pcapy, and provides an example snippet for printing packets as they arrive on an interface. The developer
    wants to see if the example works, so they create a file named snippet.py (Figure~\ref{fig:motivation:code}) containing the example code, modifying the network device name in the example, ``name of network device to capture from,'' to be ``eth0.'' However, when running \texttt{python snippet.py}, they are met with the error \texttt{ImportError: No module named pcapy} due to the fact that pcapy is not a part of the Python standard library.
    
    The developer notes that there are two packages imported in the snippet, Pcapy and Impacket. Both packages exist on the Python Package Index (PyPI), so the developer attempts to install each with \texttt{pip install pcapy impacket}. Unfortunately, Pcapy fails to install due to a compiler error. Further investigation reveals that Pcapy relies on the pcap system library headers that the developer does not have installed. The developer first attempts to install the package pcap using apt-get, their system's package manager, but no such package exists. The actual package name is libpcap0.8. However, the pcap library does not come with the headers that Pcapy requires, and the developer finds they must install the development package libpcap-dev. The final configuration is encoded by the Dockerfile in Figure~\ref{fig:motivation:Dockerfile}. Without any aid, developers face a trial-and-error struggle to discover dependencies for environment specifications such as this.
    
}

    \section{DockerizeMe}{

    The main purpose of DockerizeMe is to solve the dependency resolution problem in software configuration management. We now define the dependency resolution problem: given an runnable code snippet $C$, correctly install all \textit{language-level} and \textit{system-level} software packages required for $C$ to execute without an import error. \textit{Language-level} dependencies are dependencies managed by a package manager or tooling provided with the language runtime environment. A \textit{system-level} or \textit{system} dependency is installed on the system, but managed externally to the language runtime environment.
    
    In the context of dependency resolution, a code snippet $C$ is considered \textit{runnable} if it can be evaluated by the execution environment. That is, it does not experience fatal errors at compile or load time. A \textit{runnable} code snippet may experience a fatal error during runtime. We say $C$ experiences an \textit{import error} if it experiences a fatal runtime error caused by the failure to find a requested library.
    
    We focus on Python, a popular language with a robust ecosystem containing over 146,000 packages on its standard package platform~\cite{PyPI}. A Python code snippet experiences an import error if it exits due to Python's exception \texttt{ImportError}. We begin addressing dependency resolution by building an offline knowledge base (Sections~\ref{sec:knowledge-acquisition} and \ref{sec:knowledge-representation}). We then design an inference algorithm (Section~\ref{sec:inference-algorithm}) to return dependencies in a feasible installation order.
    
    DockerizeMe\footnote{\url{https://github.com/dockerizeme/dockerizeme}} is implemented as a NodeJS command line utility. Running \texttt{dockerizeme} on a Python package will generate the contents of a proposed Dockerfile containing all dependencies recovered by the inference procedure. 

}

    \section{Knowledge Acquisition}\label{sec:knowledge-acquisition}{

    DockerizeMe uses an offline knowledge base to correctly infer dependencies for a target script. This knowledge base contains packages, their versions and resources, and the relationships between them. It is built by applying static and dynamic analysis to known packages from the Libraries.io~\cite{LibrariesIO} dataset. Static analysis enumerates packages' known resources for later retrieval, and dynamic analysis gathers information about transitive dependencies. Association rule mining of dependencies in public Python projects takes advantage of developer generated knowledge of system level transitive dependencies. We now discuss each technique in detail.

    \subsection{Discovering Package Resources}{
    
        Inferring which packages correspond to code resources used by a script can be a challenging and non-trivial task. As reported by \cite{Horton:Gistable:2018}, many resources have a different name than the package that they belong to. It is often difficult for developers to determine which packages to use.
        
        To better inform our inference procedure, we analyzed the top ten thousand Python packages on PyPI based on their SourceRank in the Libraries.io dataset~\cite{LibrariesIO}. Packages were selected by source rank to include the most commonly used libraries, as popular libraries can affect large portions of a package ecosystem and the size of the ecosystem is prohibitive to full analysis~\cite{Hejderup:CallGraph}. If the install was successful, we recorded the distribution's top level resources as listed in \texttt{top\_level.txt}. For example, we extracted the resources Bio and BioSQL from the Python package biopython. Installation succeeded for 88\% of the tested packages.
        
        Some packages may have failed to install due to missing dependencies or some other unknown configuration. When this happened, we attempted to download and parse the package distributions manually. All packages were downloaded with pip using the options \texttt{--no-cache-dir} and \texttt{--no-deps}. If the package provided a wheel (a distribution in Python's binary format) on PyPI, we downloaded it with \texttt{--only-binary=:all:}. If the package did not have a wheel on PyPI, but did have a source distribution, we downloaded it with \texttt{--no-binary=:all:}. For source distributions, we then attempted to build a wheel distribution using the option \texttt{--no-deps}. If successful in either downloading or building a wheel for a package, we then parsed the package's top level resources by finding and reading the wheel's \texttt{top\_level.txt} file. This was successful for one in three packages.
        
    
    }
    
    \subsection{Dynamic Analysis}{
    
        Some packages may not properly list their dependencies, preventing \texttt{pip} from automatically handling resolution during install. We address this issue by performing dynamic analysis, using the 10,000  packages by SourceRank from the Libraries.io data. First, we attempt to install each package using \texttt{pip install <package>}. If the installation succeeds, we then parse the top level resources and attempt to import each. Any error output from the install/import process is logged, and on failure we parse the output for instances of the following patterns, which indicate dependence on some Python package that was not present.
        
        \begin{itemize}
            \item \texttt{no module named <name>}.
            \item \texttt{pip install <name>}.
            \item \texttt{cannot find <name>}.
            \item \texttt{cannot import name <name>}.
        \end{itemize}
        
        For example, attempting to install the Python package PyHum (\cite{PyHum}) results in the following output:
        
        \begin{quote}\itshape
            ImportError: No module named numpy. Please install numpy first, it is needed before installing PyHum.
        \end{quote}
        
        Based on the output, our dynamic analysis procedure enters a dependency record into the knowledge base which indicates that PyHum requires numpy.
    
    }
    
    \subsection{Association Rules}{
    
        Static and dynamic analysis cannot provide meaningful information about a package if the installation fails and no wheel can be found or built. Dynamic analysis may also fail to find a package's dependencies due to non-standard error messages or references to unknown header files in C libraries. In other cases, dependencies may be optional, or only required in conjunction with another package. This is the case for a simple Flask app using Raven Sentry for error logging.
        
        \begin{minted}[fontsize=\footnotesize,autogobble,breaklines,linenos,xleftmargin=15pt]{python}
        
            from flask import Flask
            from raven.contrib.flask import Sentry
            
            app = Flask(__name__)
            sentry = Sentry(app)
        
        \end{minted}
        
        Running this Flask app, after installing Flask and Raven, will result in \texttt{ImportError: No module named blinker}. The system must also have blinker, an object signaling library, installed for Raven to correctly communicate with Flask.
        
        DockerizeMe addresses these issues by augmenting its knowledge base with rules learned from existing Python environment configurations. We target public GitHub repos with a Dockerfile containing install commands for both apt and pip. The list of target repos was mined from Google BigQuery.
        
        \paragraph{Extracting items from Dockerfiles}{
        
            We inspect each repo's Dockerfile to find all \texttt{RUN} commands in both the exec and shell formats. Commands are normalized to remove new lines and escape characters. If the command is in exec format, it is additionally converted into a single command string. If a command string contains more than one command separated by \texttt{\detokenize{&&}}, \texttt{||}, or \texttt{;}, it is split into a list of individual commands.
            
            All tokens starting with ``\texttt{-}'' are assumed to be command flags and are ignored. Of the remaining tokens, commands which start with \texttt{apt-get install} are parsed for apt packages, whereas commands starting with \texttt{pip install} are parsed for pip packages. Additionally, parsed apt packages are validated to exist by checking against a list of known apt packages using the apt-cache utility. Parsed pip packages are validated to exist by checking PyPI.
        
        }
        
        \paragraph{Extracting items from requirements files}{\sloppy
        
            Some projects install Python dependencies from requirements files. Typical naming conventions are \texttt{requirements.txt} or \texttt{requirements-<env>.txt}, where env may be an environment like production or development. We look for and parse all requirements files which meet either naming convention, extracting all package specifiers within the project's requirement files using the most common subset of the requirements format specified by PEP 508. Any discovered packages are validated to exist on PyPI, and, if they exist, included in the set of pip packages.
        
        }
        
        \paragraph{Transaction format}{
        
            Parsed dependencies from each project are converted into an intermediate transaction format for association rule mining. Each project is considered as a single transaction, and its package dependencies are written as a space separated line. Package names are prefixed with the name of the package management system, either \texttt{apt\_} or \texttt{pip\_}. This is both to prevent name collisions and preserve system information while generating association rules.
            
            For example, the following Dockerfile
            \begin{minted}[fontsize=\footnotesize,autogobble,breaklines,linenos,xleftmargin=15pt]{Dockerfile}
        
                FROM python:2.7.13
                COPY snippet.py /snippet.py
                RUN ["apt-get","update"]
                RUN ["apt-get","install","-y","libmemcached-dev"]
                RUN ["pip","install","pylibmc"]
                CMD ["python","/snippet.py"]
            
            \end{minted}
            
            is parsed as the transaction
            
            \begin{minted}[fontsize=\footnotesize,autogobble,breaklines,linenos,xleftmargin=15pt]{text}
            
                apt_libmemcached-dev pip_pylibmc

            \end{minted}
        
        }
        
        \paragraph{Rule generation}{
        
            Association rules are generated from transaction data using the apriori algorithm implementation from the R package arules.\footnote{https://cran.r-project.org/web/packages/arules/arules.pdf} We use the default minimum confidence level of $0.8$. Rules are restricted to those with a maximum length of two, meaning one package in the antecedent and one package in the consequent. Minimum support is chosen to restrict to itemsets where at least three examples are seen in the transaction data. The support level was chosen to filter itemsets that occurred randomly.
        
        }
    
    }

}

    \section{Knowledge Representation}\label{sec:knowledge-representation}{

    
    We model the knowledge base as an inter-dependency graph~\cite{German:InterDependencies}, depicted in Figure~\ref{fig:dependency-graph}, stored using the Neo4J graph database. Nodes represent existing objects in the knowledge base, and directed edges represent the relationships between them. We now describe the nodes and edges used in the inter-dependency graph schema.

    \begin{figure}
    
        \centering
        
        \definecolor{package}{RGB}{230,97,1}
        \definecolor{version}{RGB}{253,184,99}
        \definecolor{resource}{RGB}{178,171,210}
        \definecolor{association}{RGB}{94,60,153}
        
        \begin{tikzpicture}[
                vertex/.style={
                    state, 
                    fill=white, 
                    drop shadow, 
                    minimum size=1.30cm, 
                    inner sep=0pt, 
                    font=\footnotesize
                },
                label/.style={
                    midway,
                    font=\footnotesize
                },
                package/.style={vertex, fill=package},
                version/.style={vertex, fill=version},
                resource/.style={vertex, fill=resource},
                association/.style={vertex, fill=association,text=white},
                edge/.style={->, line width=1.5pt}
            ]
            
            \def\delta{2}
            \def\labeldelta{0.38}
            \coordinate (origin) at ( 0, 0);
            
            \node[package]     (package)             at (origin)                         {package};
            \node[version]     (version)             at ($ (origin) + \delta*( 0, -1) $) {version};
            \node[resource]    (resource)            at ($ (origin) + \delta*(-1,  0) $) {resource};
            \node[association] (association)         at ($ (origin) + \delta*( 1,  0) $) {association};
            \node[package]     (consequent)          at ($ (origin) + \delta*( 1, -1) $) {package};
            \node[resource]    (dependency-resource) at ($ (origin) + \delta*(-1, -2) $) {resource};
            \node[version]     (dependency-version)  at ($ (origin) + \delta*( 0, -2) $) {version};
            \node[package]     (dependency)          at ($ (origin) + \delta*( 1, -2) $) {package};
            
            \draw[edge] (package)            to              node [label, left]              {version}              (version);
            \draw[edge] (version)            to [bend left]  node [label, below left]        {resource}             (resource);
            \draw[edge] (package)            to              node [label, above=\labeldelta] {association}          (association);
            \draw[edge] (association)        to              node [label, left]              {association}          (consequent);
            \draw[edge] (version)            to [bend right] node [label, below right]       {resource\_dependency} (dependency-resource);
            \draw[edge] (dependency-version) to              node [label, above=\labeldelta] {resource}             (dependency-resource);
            \draw[edge] (dependency)         to              node [label, above=\labeldelta] {version}              (dependency-version);
            
        \end{tikzpicture}
        \caption{
            Relationships between the resource nodes in the DockerizeMe inter-dependency graph.
        }
        \label{fig:dependency-graph}
        
    \end{figure}
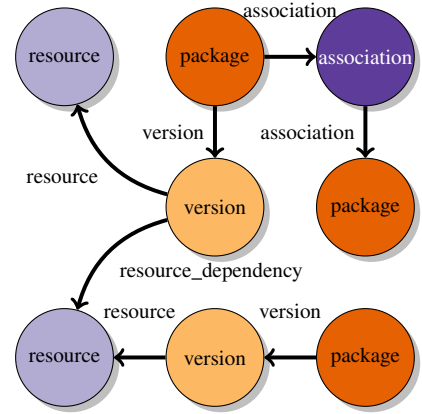
    
    \paragraph{Package Nodes}{
    
        Each unique package known to DockerizeMe is stored as a node in the inter-dependency graph. Package nodes are tagged with the label \texttt{package} and store both the package's name and package management system. We enforce that the name and package management system be unique together. That is, no two packages served by the same system may have the same name.
    
    }
    
    \paragraph{Version Nodes}{
    
        All known versions of a package are represented as a version node. Versions are tagged with the label \texttt{version} and store the package version number. Additionally, a directed \texttt{version} edge connects the package to its version.
    
    }
    
    \paragraph{Resource Nodes}{
    
        A resource is one of the directly importable package objects discovered during static analysis. Nodes are tagged with the label \texttt{resource} and store the object name. Because a package's resources may change between versions, resource nodes are owned by version nodes. This is indicated by a directed edge from the version node.
    
    }
    
    \paragraph{Modeling Dependencies}{
    
        Dynamic analysis discovers a package version's dependencies on external resources. We model this relationship in the knowledge base as a directed \texttt{resource\_dependency} edge from the version node to a resource node matching the name of the required resource.
    
    }
    
    \paragraph{Association Nodes}{
    
        Association nodes represent individual association rules. Nodes are tagged with \texttt{association} and maintain metadata for \texttt{confidence}, \texttt{support}, \texttt{lift}, and \texttt{count}. Directed \texttt{association} edges connect packages with their association rules. A \texttt{package $\rightarrow$ association} edge means that the package is in the rule antecedent, and an \texttt{association $\rightarrow$ package} edge means the package is in the association rule consequent.
    
    }

}

    \section{Inference Algorithm}\label{sec:inference-algorithm}{

    When given a target application, the goal of inference is to determine all dependencies required to run the application without error. There is an additional constraint that dependencies must be returned in a correct order. For example, if some package A depends on B and C at install time and B also depends on C at install time, then the correct order of installation must be C, B, A.
    
    The inference algorithm first extracts imported resources from the target application (Section~\ref{subsec:parsing-target-code}) and interrogates the knowledge base to determine the set of packages that the resource likely belongs to (Section~\ref{subsec:reverse-lookup}). The inter-dependency graph is then traversed starting at each of these root nodes to determine transitive dependencies (Section~\ref{subsec:dfs}).
    
    \subsection{Parsing Target Code}\label{subsec:parsing-target-code}{
        
        The first task is to parse a target application and extract a list of all imported resources. We do this by building an abstract syntax tree (AST) of the source code. In Python, imported resources are defined by \texttt{Import} and \texttt{ImportFrom} nodes, which correspond to the statements \texttt{import <package>} and \texttt{from <package> import <resource>}. The full resource name at an \texttt{Import} node is the node name. For \texttt{ImportFrom} nodes, the full resource name is the name of the module plus the names of each resource being imported. Algorithm~\ref{alg:visiting-imports} outlines pseudocode for visiting both node types in the AST.
        
        \begin{algorithm}
        
            \Procedure{VisitImport(node)}{}{
                
                \If{not IsStandardLibrary(node)}{
                
                    import\_libraries += name(node)\;
                    
                }
            
            }
            
            \Procedure{VisitImportFrom(node)}{}{

                \If{not IsStandardLibrary(node)}{
                
                    \For{resource in resources(node)}{
                    
                        import\_libraries += (\;
                        \Indp 
                            name(module(node)) + name(resource)\;
                        \Indm
                        )\;
                    
                    }
                
                }
            
            }

            \caption{
                Procedures for parsing imported libraries from Import and ImportFrom node types in a Python AST.
            }
            \label{alg:visiting-imports}

        \end{algorithm}
        
        Resources are filtered to exclude those in the standard library. We perform filtering by checking to see if the resource exists in a clean Python environment with Python import tools located in the \texttt{imp} module. Pseudocode is provided in Algorithm~\ref{alg:is-std}. First, we look up the file system path of a resource by the resource name. If path lookup fails with an \texttt{ImportError}, we know that the module cannot possibly be part of the standard library. If lookup succeeded, we verify that the resource name either matches that of a known Python builtin or that the resource path does not contain \texttt{site-packages} and \texttt{Extras}. Extra packages are sometimes included with a Python distribution, but are not a part of the standard library, and \texttt{site-packages} is where pip places other installed packages by default.

        \begin{algorithm}
            
            \Procedure{IsStandardLibrary(module)}{

                path = getPathOrError(module)\;
                \KwRet isBuiltin(module) or (\;
                \Indp \textquotesingle{}site-packages\textquotesingle{} not in path\;
                    ~and \textquotesingle{}Extras\textquotesingle{} not in path\;
                \Indm)\;
                
            }
            
            \caption{
                A Python module is considered part of the standard library if it is a builtin or it isn't installed in a location reserved for third-party packages.
            }
            \label{alg:is-std}
            
        \end{algorithm}
    
    }
    
    \subsection{Mapping Resources to Packages}\label{subsec:reverse-lookup}{
    
        Once the resources for an application are known, they must be mapped back to a set of packages that can be installed. We perform this reverse lookup by querying our knowledge base and the package management system of record for potential matches. A match between a resource required by an application and an installable package may be determined by a full or partial match on one or more known resources in the knowledge base, or a full match on a known package (either in the database or through the package management system).
        
        \begin{algorithm}
        
            \Procedure{MapResourcesToPackages(resources)}{
            
                packages = set()\;
                
                \For{resource in resources}{
                
                    packages += queryKnownResourcesStartingWith(\;
                        \Indp\Indp\Indp\Indp name(resource))\; \Indm\Indm\Indm\Indm
                    packages += queryKnownPackagesByName(\;
                        \Indp\Indp\Indp\Indp name(resource))\; \Indm\Indm\Indm\Indm
                        
                    \If{not containsByName(packages, name(resource))}{
                    
                        packages += queryPackageManagementSystem(\;
                        \Indp\Indp\Indp\Indp name(resource))\; \Indm\Indm\Indm\Indm
                    
                    }
                
                }
                
                \KwRet packages\;
            
            }
            \caption{
                Mapping a list of resources imported by a code snippet to a list of packages to which they may belong.
            }
            \label{alg:resources-to-packages.}
        
        \end{algorithm}
        
        A full match on a resource queries the knowledge base for any known resources whose names exactly match the name of a resource used by the application. If any such resources are found, the packages that own them are returned. We filter package results to be distinct, as it may be the case that some package has multiple versions with the same resource.
        
        Partial matches search for any known resources where the name of a resource used by the application starts with the name of the of the resource in the knowledge base. For example, consider the simple Python script below:
        
        \begin{minted}[fontsize=\footnotesize,autogobble,breaklines,linenos,xleftmargin=15pt]{Python}
            import zope.interface

            class Interface(zope.interface.Interface):
                attr = zope.interface.Attribute('Attribute')
            
            print(type(Interface))
        \end{minted}
        
        The script imports the resource \texttt{zope.interface}, which corresponds to a package on PyPI by the same name. However, the \texttt{zope.interface} package has a top level resource of the name \texttt{zope} with a submodule named \texttt{interface}. The partial match ensures that packages whose resources follow this naming convention get matched when performing reverse lookup. Looking for exact matches is a special case of partial matches where the full resource names are equivalent.
        
        Additionally, we check to see if any package exists with the same name as a required resource. Previous work showed that this happens approximately 45\% of the time, and doing so may result in finding the correct package for a resource even if its resources could not be discovered through static analysis~\cite{Horton:Gistable:2018}.

        When reverse lookup has completed, package names are normalized to match the name on the package management system, as some systems treat certain characters as being identical. For example, according to PEP 508~\cite{PEP:508}, Python does not distinguish between a dash and an underscore, so \texttt{flask\_heroku} matches the package \texttt{flask-heroku}.

    }
    
    \subsection{Transitive Dependency Recovery}\label{subsec:dfs}{
    
        Knowing only the packages corresponding to top level resources is often not sufficient for correct environment configuration, as those packages may themselves depend on other packages. Information about the transitive dependency of packages is encoded in the inter-dependency graph through \texttt{resource\_dependency} and \texttt{association} relationships. Assuming the inter-dependency graph contains all necessary relations, the set $P$ of packages which must be installed is the set $S$ of resolved direct dependencies (Section~\ref{subsec:reverse-lookup}) unioned with the set $R$ of packages reachable from $S$.
        
        It is, however, not sufficient just to compute $P$. We must also preserve a correct ordering of dependencies such that each package is installed before any other package which depends on it. We do this by performing a depth first search rooted from each package $p \in S$, where neighboring packages are computed by following the directed \texttt{resource\_dependency} and \texttt{association} relationships. An ordered list of packages is maintained throughout the DFS, and a package $p$ is added to the list once all of its children have been traversed. As in our lookup procedure, the names of each package are normalized using the specific package management system they belong to.
        
        In an acyclic graph, the ordering returned by our DFS based transitive dependency resolution results in a reverse topological order. However, we cannot guarantee that our inter-dependency graph is acyclic. It may be the case that two packages (either correctly or incorrectly) depend on each other. Additionally, if two packages $p_1$ and $p_2$ are frequently used together, our association rule mining may have generated the rules $p_1 \rightarrow p_2$ and $p_2 \rightarrow p_1$. We use our DFS resolution as a heuristic.
    
    }

    \begin{algorithm}
    
        \Procedure{RecoverTransitiveDependencies(packages)}{
        
            encounteredPackages = set()\;
            dependencies = list()\;
            
            \Procedure{DFS(node)}{
                
                \If{node in encounteredPackages}{ \KwRet\; }
                
                encounteredPackages += node\;
                
                directDependencies = (\;
                \Indp queryPackageDependencies(node)\; \Indm
                )\;
                \For{dependency in directDependencies}{
                
                    DFS(dependency)\;
                
                }
                
                dependencies += node\;
            
            }
            
            \While{root = first(packages)}{
            
                DFS(root)\;
            
            }
        
            \KwRet dependencies\;
        
        }

        \caption{
            Depth-first search used to discover transitive dependencies in a correct order.
        }
        \label{alg:transitive-recovery}
    
    \end{algorithm}
    
}

    \section{Evaluation}{

    Gistable (\cite{Horton:Gistable:2018}) introduces the Gistable dataset, a collection of 10,259 single-file Python scripts mined from GitHub's public gist service. Gists in the dataset were discovered by querying the GitHub gist search UI for gists in the Python language with at least one star rating and automatically scraping the returned results. Analysis of the Gistable dataset showed that the most common exit status of gists in the dataset was an import error, occurring 52\% of the time. In particular, the import errors in 2,891 gists could not be fixed by Gistable's naive approach of attempting to install a package named for each of the imported resources not in the standard library. We focus on these ``hard'' gists for which Gistable's naive inference algorithm was not sufficient, referring to them as \hardgists{}.

    \subsection{Methodology}{

        To evaluate DockerizeMe, we analyzed its ability to remove import errors from gists in \hardgists{}. The first step in the analysis of each gist was to use DockerizeMe to perform dependency resolution to generate a list of packages $P$ to install. We then installed each package $p \in P$ using the Python package manager pip in a clean Python 2.7.14. The execution result of each gist was recorded as the name of the exception raised when run, if any, or \texttt{Success}. We consider a gist to be fixed in the context of dependency resolution if its exit status is anything other than \texttt{ImportError}.

        Installation failures encountered while evaluating a gist were ignored. The rationale for this choice being that if the failed package is a direct or transitive dependency of the gist, execution will raise \texttt{ImportError}. However, the package may not necessarily be a true dependency, due to the nature of the association rules in the DockerizeMe knowledge base. If so, it is acceptable to continue without the package.

        All gists were evaluated in a Docker container based off of the \texttt{python:2.7.14} image. The DockerizeMe Docker image configured the aptitude package manager to go through a local proxy, for efficiency, and configured pip by disabling filesystem cache and setting the default timeout to 10 minutes. Analysis jobs were scheduled on a Nomad cluster, a process and container management system that natively supports scheduling Docker containers. Compute nodes were running Ubuntu 16.04 with 4 CPUs and 8GB memory.

    }

    \subsection{Results}{

        We rank the quality of an inferred environment primarily by whether or not its corresponding gist experiences an import error during execution. For successful inferences, we also consider the characteristics of the inferred environment by number of direct and transitive dependencies and the total number of dependencies overall. We then address installation failures encountered in inferred environment configurations and reasons why the inference procedure may have failed to produce a working environment. Finally, we consider the exit status of fixed gists as an indication of future issues to address in automated environment configuration.

        \subsubsection{Inference}{

            We evaluated DockerizeMe's inference procedure on a total of 2,891 Python gists from \hardgists{}, a subset of the Gistable dataset chosen as a baseline that cannot be fixed by installing each package by its resource name, the natural first step taken by developers when attempting to resolve import errors~\cite{Horton:Gistable:2018}. Our evaluation found that, of the environments generated by installing the dependencies discovered by DockerizeMe's inference procedure, an additional 892 gists (31\%) executed without experiencing Python's \texttt{ImportError}. This result demonstrates an important step in automated inference of environment configuration. We discuss future improvements in the next section below.

        }

        \subsubsection{Environment Characteristics}{

            On average, the gists from \hardgists{} for which DockerizeMe generated a working environment specification import 2{\textendash}3 unique resources, consistent with the overall average for all gists in \hardgists{}. DockerizeMe reported performing at least one name resolution for the top 79\% of the gists, meaning that most gists import a resource which was mapped to a package with a different name. 40\% had at least one transitive dependency. Overall, imported resources were mapped to an average of three direct dependencies with an additional two transitive dependencies being found by the inference procedure. Transitive dependencies were most likely to be additional Python packages.

        }

        \subsubsection{Size}{

            The largest environment specification proposed by DockerizeMe contained 206 unique Python and APT packages. Python packages held a larger share of the total number of packages installed, with a maximum of 195 being installed in an environment versus 48 APT packages, respectively. Comparatively, the largest environment specification among those that fixed their gist's import errors only contained 87 unique packages, with the maximum of 80 Python package installed and 9 APT packages being installed in any environment.

            In both cases, the large environment size was an outlier, with most working configurations being over $17\times$ smaller. The size of the largest working configurations suggests the need for a post-inference reduction process capable of reducing a set of proposed packages to a minimal set which still resolve a script's import errors.

        }

        \subsubsection{Frequency of Installation Failures}{

            Our evaluation procedure ignored installation errors for inferred dependencies under the assumption that if a package failed to install, but the overall configuration still worked, the failed package can be removed from the final configuration. Only 9 of the 892 fixed gists (1\%) experienced an installation error while building the inferred environment but still exited without experiencing an import error. Manual inspection revealed that 4 of the failures were due to an unknown dependence on a Python package and 1 was due to an unknown dependence on a system package. The remainder failed either due to issues with the packages themselves or due to network noise.

        }

        \subsubsection{Remaining Reasons for Failure}{

            1,999 gists still exited with Python's \texttt{ImportError} after applying inference to all gists in \hardgists{}. We performed an additional qualitative coding process on a random sample of 30 such gists, along with the generated environment specification, to determine why inference failed to generate a working environment.

            While performing coding, each gist was inspected to determine the root cause of its exit status. We then reran DockerizeMe's inference algorithm, inspecting the inference process at runtime to determine why the root cause was not repaired. We coded each gist according to the ultimate reason why a correct environment specification was not generated. We employed negotiated agreement during the coding process to address the reliability of coding~\cite{Campbell2013}. Using this technique, the first and second authors work collaboratively to clarify definitions of codes and reach agreement on the assigned code.

            The most common reason for failure, occurring for 15 out of the 30 gists analyzed, was that the environment specification did not find a direct dependency. That is, the gist imported some resource that was not present after installing all inferred dependencies, indicating that there was some mapping from a resource to a Python package that DockerizeMe did not know about. Four gists failed because environment inference did not find a transitive dependency, meaning that an inferred dependency failed to install because of additional configuration requirements or that execution in the inferred environment resulted in an import error for a resource not directly imported by the gist. Another three gists required had the correct package inferred for their dependencies, but needed another version due to breaking changes made to the package's API.

            The remainder of the gists failed with import error, but could not have been fixed by inferring dependencies from PyPI or APT. For example, some gists required an execution environment other than the system CPython interpreter. One such gist was written as a Sublime Text plugin and required being run from within the editor. Other gists imported local configuration files that were meant to be written by the user.

            Mapping resource names to their corresponding packages remains the largest issue likely due to the size of the packaging ecosystem (DockerizeMe analyzed only the top 6\% of packages and 1\% of versions on PyPI when generating the knowledge base). A straightforward, though expensive, solution to the mapping problem already exists: pre-process the remaining package versions. However, given the large and continually growing environment, environment inference procedures can benefit from improved heuristics for predicting what packages may belong to a resource without prior indexing.

        }

        \subsubsection{Exit Status of Fixed Gists}{

            \begin{table}[ht]
                \centering
                \caption{
                    Exit status of the 892 gists from \hardgists{} for which DockerizeMe's inference algorithm was capable of repairing import errors, filtered to those with a count greater than 20.
                }
                \label{tbl:fixed-exit-status}
                \begin{tabular}{lrr}
                    \toprule

                    \textbf{Exit Status} & \textbf{Count} & \textbf{Percent} \\

                    \midrule

                    Success & 473 & 53.0\% \\
                    NameError & 145 & 16.3\% \\
                    ImproperlyConfigured & 91 & 10.2\% \\
                    IOError & 41 & 4.6\% \\
                    SystemExit & 23 & 2.6\% \\
                    AttributeError & 23 & 2.6\% \\
                    RuntimeError & 23 & 2.6\% \\
                    \textit{Other} & 73 & 8.1\% \\

                    \bottomrule
                \end{tabular}
            \end{table}

            Table~\ref{tbl:fixed-exit-status} shows the exit status of gists from \hardgists{} which had their import errors resolved by DockerizeMe. The most common exit status is \texttt{Success}, occurring for over 50\% of the gists. The next most common exit status was Python's \texttt{NameError}, meaning that the gist attempted to access an object reference which did not exist. In most cases, instances of \texttt{NameError} are an issue with the gist itself and outside the focus of DockerizeMe. An exception to this is the use of wildcard imports where the resources provided by a package have changed. For future research, we can determine with static analysis if this use case is common and in need of addressing.

            Of the remaining 30\%, 91 exited due to Django's \texttt{ImproperlyConfigured} error and 41 due to \texttt{IOError}. \texttt{ImproperlyConfigured} is an exception raised by the Django framework on initialization if it is missing a required configuration in the settings file. \texttt{IOError} can indicate an issue communicating with a service, such as a database server.

            The most common exit codes after applying inference closely resemble those found by Gistable (\cite{Horton:Gistable:2018} Table I), with the exception of \texttt{SyntaxError} and its child \texttt{IndentationError}, which were filtered out by the selection criteria.

            Future research can evaluate how gists rely on configuration files, environment variables, and external services with the intention of generating an environment configuration that provides these resources.


        }

        \subsubsection{Summary}{

            We conclude with a summary of our results.

            \vspace{0.5em}
            \noindent{\insight{
                Our inference algorithm recovered an additional 31\% of environment configurations over the baseline approach. DockerizeMe was able to successfully resolve the correct packages when there was no direct name match, and discover transitive dependencies. Python gists often require non-trivial environment configuration in order to run. Future work is needed to handle other configuration steps, such as missing environment variables and providing expected services.
            }}

        }

    }

    \subsection{Inference on Real-World Snippets}{

        The following snippets illustrate how DockerizeMe's inference algorithm overcomes challenges to provide a correct environment specification.

        \subsubsection{Package name resolution}{

            The first challenge to overcome is determining which packages provide the resources used by some code snippet. As \cite{Horton:Gistable:2018} discovered, determining the correct package to install is still a challenging task for developers with experience in performing software configuration.

            Consider the following snippet, which imports the resource PIL and uses it to create and save a new image.

            \begin{minted}[fontsize=\footnotesize,autogobble,breaklines,linenos,xleftmargin=15pt]{Python}

                from PIL import Image
                img = Image.new('RGB', (100, 100))
                img.save('image.png')

            \end{minted}

            PIL is not a part of the Python standard library, so the package containing it must be installed. There \emph{is} a package named PIL on PyPI. However, attempting to install it reveals that the package is not maintained and has no published versions that can actually be installed.

            DockerizeMe realizes that the package PIL cannot be installed because it has no available versions. Further, it knows that another Python package, Pillow, also has a resource named PIL. The inference algorithm recommends the following Dockerfile which runs the snippet without import error.

            \begin{minted}[fontsize=\footnotesize,autogobble,breaklines,linenos,xleftmargin=15pt]{Dockerfile}

                FROM python:2.7.13
                COPY snippet.py /snippet.py
                RUN ["pip","install","Pillow"]
                CMD ["python","/snippet.py"]

            \end{minted}

        }

        \subsubsection{Transitive dependencies}{

            Another challenge faced by dependency resolution is determining transitive dependencies. In some cases, a snippet can rely on a package which itself relies on one or more other packages. The following snippet demonstrates this using the module dashtable to convert an html formatted table to a GitHub markdown table.

            \begin{minted}[fontsize=\footnotesize,autogobble,breaklines,linenos,xleftmargin=15pt]{Python}

                import dashtable
                print(dashtable.html2md("""
                    <table>
                        <tr><th>Header 1</th><th>Header 2</th></tr>
                        <tr><td>Data 1</td><td>Data 2</td></tr>
                    </table>
                """))

            \end{minted}

            There exists a package by the name of dashtable on PyPI, and it can be installed. However, running the snippet after installing dashtable results in the error \texttt{ImportError: No module named bs4}. This is because dashtable relies on the module bs4 to parse html. Fortunately, dashtable is in DockerizeMe's knowledge base, and the inference algorithm correctly infers that dashtable relies on beautifulsoup4, the package that provides the module bs4. Running the snippet with the DockerizeMe generated dockerfile correctly results in printing the converted table.

            Often, a snippet may have a transitive dependency on a package which is not served over PyPI. Consider the following snippet, which makes use of the module pylibmc. The pylibmc package on PyPI fails to compile in a clean environment because it is missing the header file \texttt{memcached.h}.

            \begin{minted}[fontsize=\footnotesize,autogobble,breaklines,linenos,xleftmargin=15pt]{Python}

                import pylibmc
                mc = pylibmc.Client(["127.0.0.1"])

            \end{minted}

            Association rule mining provides the correct inference here. In the configuration scripts that were parsed when building the knowledge base, any script that installed pylibmc was also likely to install libmemcached-dev using the apt-get package manager. DockerizeMe proposes a Dockerfile which installs memcached before installing pylibmc, allowing the snippet to be executed without error.

        }

    }

    \subsection{Limitations}{

        While our technique can discover unspecified dependencies, there are limits to the types of inference that can be performed with our current knowledge sources. We now present each limitation and the reason behind them.

        \paragraph{Incomplete Knowledge}{

            DockerizeMe's knowledge base of Python packages was built by analyzing the top 10 thousand packages as listed in the Libraries.io dataset, sorted by SourceRank. Association rule mining of packages found in Dockerfiles and requirements files of public repos on GitHub was responsible for populating known apt packages.

            While our process for knowledge generation is designed to target the most frequently used packages, it does not have complete knowledge of the ecosystem. The apt ecosystem has over 42 thousand packages in the default repo. PyPI has almost 150 thousand packages with over 1 million unique versions available for download.

            While analyzing the full ecosystem would resolve inference failure due to not knowing about a specific dependency, such analysis would be difficult to complete within a feasible time frame. In addition, brute forcing public registries cannot inform the knowledge base about packages that are available through git or are hosted on private mirrors.

        }

        \paragraph{Version Inference}{

            Knowledge base generation and the subsequent inference procedure only take into account the latest version of a package. There are, however, reasons why a code snippet may depend on another version of a package, including deprecation, removal, and renaming. Additionally, snippets and dependencies requiring a different Python version are unsupported.

        }

        \paragraph{Package Oriented}{

            DockerizeMe focuses on enabling execution of code snippets with import errors by inferring their software dependencies. When this fixes import errors, the exit status is considered to be \texttt{Success}. However, the gist may have a dependency on a running service. For example, a gist which imports the database drivers for MySQL will likely attempt to connect to a MySQL server. While DockerizeMe can install the database package, it currently cannot configure and run the database service.

        }

        \paragraph{Non-Installable Packages}{

            Some packages available in the Python and apt ecosystems may not be capable of being installed. For example, Python Quartz had a known bug that caused a fatal error on installation due to an improper configuration with the packaging system~\cite{Quartz:Wontfix}. Another package, PyTorch, was not available for install through PyPI, but the maintainers hosted an empty package on PyPI which failed installation with an informative message for the user~\cite{PyTorch:NotHosted}.

            DockerizeMe cannot produce a correct inference under either circumstance, as it assumes that packages discovered during knowledge acquisition are installable.

        }

        \paragraph{Hardware and OS Requirements}{

            Some packages require specific hardware, like a Raspberry Pi, or a specific operating system, like Windows. DockerizeMe cannot provide a hardware configuration, and does not take into account the current configuration of the system when building a Dockerfile. In addition, the DockerizeMe knowledge base was built with system dependencies from the apt package ecosystem, and has no knowledge of system dependencies for Windows, which does not have an official package manager.

        }

    }

}

    \section{Discussion and Future Work}{

    Although DockerizeMe's inference algorithm is capable of fixing import errors encountered in nearly a third of \hardgists{}, future work is needed to address import errors in gists which still fail with \texttt{ImportError} after applying our inference procedure. Other techniques are needed for improving the quality of an inferred environment, either by improving the inference algorithm or as an additional post-processing step.

    \paragraph{Exploration of Other Knowledge Sources}{
    
        DockerizeMe's knowledge base is generated from the results of static and dynamic analysis of public Python packages, adding associations from publicly available Dockerfiles. These are, however, not the only knowledge sources available for use. Other potential knowledge sources include additional build or configuration scripts such as Vagrantfiles and continuous integration logs from services like TravisCI.
        Static and dynamic analysis using system dependence graph techniques~\cite{Liang:SlicingObjects, Sinha:DependencyGraphSlicing} may be performed to enable inference at the package version level by detecting reliance on deprecated features or breaking changes in a package's API.
        
        Additionally, developer generated knowledge present on sites like Stack Overflow may be usable data sources. It may be sufficient to parse accepted answers for code blocks indicating a list of packages to install, using such lists as transactions for association rule mining. Another area of research may investigate using natural language processing to parse questions related to a problem encountered during inference, then automatically apply suggested solutions in highly rated answers.
        
    }

    \paragraph{Breaking Cycles}{

        We currently use DFS to return a reverse topological ordering as a heuristic, as it is only guaranteed to be correct so long as the underlying graph is acyclic. Due to the nature of the packaging ecosystem and association rule generation, we cannot guarantee that directed cycles do not occur in the inter-dependency graph.

        Future research will need to focus on the prevalence of such cases, and whether or not they are responsible for issues in the build process. In cases where assumptions in our DFS implementation result in an incorrect installation order for dependencies, we can try other methods of breaking cycles. One method might be to compute the feedback edge set of each connected component to guarantee the minimum number of back-edges are disregarded.

    }

    \paragraph{Feedback Directed Inference}{

        A majority of the inference process is performed offline. That is, knowledge acquisition occurs prior to inference, and the inference procedure itself generates a single static configuration. This does not give inference the ability to backtrack or recover from errors in cases where an incorrect dependency is resolved, except for the naive process of ignoring installation errors.

        Future research will focus on feedback directed inference, a process where inference is performed iteratively in tandem with analysis. Performing inference online will allow DockerizeMe to determine if applying a configuration results in the resolution of an import error. If it does not, inference may backtrack and attempt a different configuration instead.

        Applying feedback directed inference may help improve the overall quality of environment specifications, reduce the need for post-inference minimization, and allow the inference procedure to learn as it processes gists.

    }
    
    \paragraph{Post-processing to reduce unnecessary dependencies}{
    
        The largest working configuration inferred by DockerizeMe for \hardgists{} contained 87 unique packages, while most working configurations were over $17\times$ smaller. This suggests the need for an inference method to reduce an environment specification to a minimal working configuration. 
        
        Delta debugging may be a viable method for minimizing the environment specification. Dockerfiles produced by DockerizeMe install dependencies with a dedicated \texttt{RUN} instruction. Starting with a complete specification as produced by DockerizeMe, we can delta debug by removing \texttt{RUN} instructions until a minimal set of dependencies remain.

        The challenge with delta debugging Dockerfiles, in contrast to delta debugging in a standard application repair context, is that there may not be an easily executable test suite to evaluate the quality of the environment. Even in the context of DockerizeMe, where the success criteria is informed by whether or not the executable exits with a particular status, the time necessary to build a complete environment can be prohibitive. Many Python dependencies, in addition to the time required to download, also require being compiled against header files in the local filesystem. The extra compilation step adds to the overall build time.
        
        Docker does have the ability to cache layers for every independent stage in the build. However, the cache is only valid if the parent layer is already in the cache and the instruction string exactly matches one used to generate a child layer in the cache. This restriction greatly reduces the ability to leverage the build cache during delta debugging. It may be possible to determine the optimal order for delta debugging to maximize cache usage. We may also be able to exploit sparsity in DockerizeMe's inter-dependency graph and perform delta debugging over independent components. If independent components are small on average relative to the size of the inferred environment, the total size of the search space can be greatly reduced.

    }
    
    \paragraph{Additional Languages}{
    
        DockerizeMe focuses on Python, but its inference procedure and mining procedures only require that source code can be parsed for imported and exported resources. We believe our approach generalizes to languages like R or NodeJS, because such languages meet our requirements. Future research can assess DockerizeMe's ability to handle configuration for such languages.
    
    }

}

    \section{Related Work}{
    
    Cito et al. investigated the current state of the Docker ecosystem by inspecting public GitHub repositories. They found that only 66\% of public Dockerfiles can be built, with most quality issues being caused by dependency issues and most Dockerfile changes being made to address build dependencies~\cite{Cito:DockerEcosystem}. The most common dependency error, according to their analysis, was the failure to pin a dependency version. While their study focuses on the buildability of Docker containers, it remains an empirical analysis and makes no attempt to repair broken configurations.
    
    Work by Hassan and Wang has focused directly on repairing configuration build scripts. HireBuild (History-Driven Repair of Build Scripts), is designed to repair failing gradle build scripts based on potential repairs discovered from the TravisTorrent dataset~\cite{Hassan:HireBuild}. Their work is similar to ours in that it incorporates knowledge from existing, developer driven sources and uses that information to infer a correct environment specification codified in a build script. Other approaches to configuration repair target inconsistencies between file systems and configuration scripts~\cite{Su:autobash:2007} or capture and replay developer changes~\cite{Weiss:Tortoise:2017}. Macho et al. focus on repairing Java projects using the Maven build system by updating dependency versions, deleting listed dependencies, or explicitly specifying common repositories~\cite{Macho:2018:BuildMedic}.
    
    In contrast to these approaches, our work focuses on generating a complete environment specification for a codebase without a prior configuration or developer input. To our knowledge, this is the first successful use of analysis to infer environment specifications entirely from scratch.
    
    Related to our proposed future investigation of applying delta debugging to find a minimal environment specification is Cimplifier, a technique for partitioning a container's resources into a set of smaller, independent containers~\cite{Rastogi:Cimplifier}. A natural side effect of the Cimplifier process is container slimming due to only maintaining the resources which are used during application execution. A similar process may be a reasonable alternative to delta debugging, but may suffer from the limitations of dynamic analysis to uncover all dependencies.

}

    \section{Conclusion}{

    We investigate a technique for the automatic inference and configuration of a computing environment capable of executing an arbitrary Python code snippet without resulting in an import error. Our technique builds a knowledge base of known Python packages. It discovers information about package dependencies using a combination of static analysis, dynamic analysis, and association rule mining. Dependencies are resolved in a correct order for installation. Finally, we provide a tool, DockerizeMe, which delivers an environment configuration as a Dockerfile for the Docker container system. Results from our study showed that DockerizeMe resulted in a 30\% improvement over a baseline approach to environment configuration.

    DockerizeMe is a fist step in automating environment configuration, a process with the potential to save developer time and effort, as well as reducing potential mistakes in application deployments. While we focus on enabling the executability of Python code snippets, we believe our inference procedure can extend to other languages with a package management ecosystem. Future research will focus on improving the inference procedure and investigate support for other configuration properties such as environment variables or external services.

    {\footnotesize

        \textbf{Acknowledgments}:
        This work is funded in part by the NSF SHF grant \#1814798. We would like to thank Kai Presler-Marshall for his help parsing Stack Overflow data and validating packages in the aptitude ecosystem.

    }

}

    \bibliographystyle{IEEEtran}
    \bibliography{main}

\end{document}